\documentclass[prb,twocolumn]{revtex4}
\usepackage{graphicx}

\begin{document}

\title{Pressure-induced insulator-to-metal transition in low-dimensional TiOCl}

\author{C. A. Kuntscher,$^{1*}$ S. Frank,$^{1}$ A. Pashkin,$^{1}$ M. Hoinkis,$^{2,3}$
M. Klemm,$^2$ M. Sing,$^3$ S. Horn,$^2$ and R. Claessen$^3$}
\address{
$^1$ 1. Physikalisches Institut, Universit\"at Stuttgart, Pfaffenwaldring 57, D-70550 Stuttgart, Germany\\
$^2$ Experimentalphysik 2, Universit\"at Augsburg, D-86135 Augsburg, Germany \\
$^3$ Experimentelle Physik 4, Universit\"at W\"urzburg, D-97074 W\"urzburg, Germany}

\date{\today}

\begin{abstract}
We studied the transmittance and reflectance
of the low-dimensional Mott-Hubbard insulator TiOCl in the infrared and visible
frequency range as a function of pressure. The strong suppression of the
transmittance and the abrupt increase of the near-infrared reflectance above
12 \nolinebreak GPa suggest a pressure-induced insulator-to-metal transition.
The pressure-dependent frequency shifts of the orbital excitations, as well as the
pressure dependences of the charge gap and the spectral weight of the optical
conductivity above the phase transition are presented.
\end{abstract}


\maketitle
Low-dimensional titanium-oxychloride, TiOCl, is under discussion recently as a spin-Peierls
system with puzzling properties, like for example the occurrence of two successive
phase transitions.\cite{Seidel03,Imai03,Kataev03,Hemberger05}
The first-order phase transition
at T$_{c1}$=67 K was attributed to a spin-Peierls transition with a dimerization of
the chains of Ti atoms along the $b$ axis.\cite{Shaz05} The nature of the second-order phase transition at
T$_{c2}$=91 K is still under discussion; a transition to an incommensurate spin-Peierls state below
T$_{c2}$ with a subsequent lock-in transition to a commensurate dimerized state below
T$_{c1}$ was proposed,\cite{Krimmel05,Smaalen05,Ruckkamp05a} but also strong orbital 
fluctuations with a near degeneracy of the lowest-lying orbitals were considered.
\cite{Imai03,Kataev03,Hemberger05,Saha04,Lemmens04}

The importance of the orbital degree of freedom in TiOCl was recently addressed
by polarization-dependent transmission measurements in the infrared and
visible frequency range.\cite{Ruckkamp05a,Ruckkamp05b} In the transmittance spectra
absorption features were observed at 0.6-0.7 eV for {\bf E}$||$$a$ and
at 1.3-1.6 eV for {\bf E}$||$$b$, which were interpreted in terms of 
excitations between the crystal field-split Ti $3d$ energy levels.
According to these results, the orbital degree of freedom is quenched
due to a significant crystal field splitting of the $t_{2g}$ orbitals.
It was therefore concluded that the interesting physics of TiOCl
can be entirely explained by the interplay of the lattice and spin degrees of
freedom.\cite{Ruckkamp05a,Ruckkamp05b}

TiOCl is a Mott-Hubbard insulator: Each Ti $3d$ shell is occupied
by one electron, and due to electronic correlations these charge
carriers are localized on-site. However, the carrier localization
effects should be relatively weak because of the high
nearest-neighbor exchange coupling.\cite{Seidel03,Kataev03} It was
even suggested that TiOCl is close to an insulator-to-metal
transition.\cite{Seidel03,Beynon93,Craco04} This makes the
material an interesting candidate for doping, leading to a
metallic and possibly superconducting state,\cite{Seidel03,Craco04} like,
for example, observed in low-dimensional
$\beta$-Na$_{0.33}$V$_2$O$_5$ under pressure.\cite{Yamauchi02}
Unfortunately, up to now doping of TiOCl could not be achieved.

Instead of such a band{\it filling} control of the insulator-to-metal transition
one can control the {\it width} of the energy bands close to the Fermi energy.\cite{Imada98}
This can be accomplished by applying high external pressure.
We followed this route in our study whose results we present here.
In our optical experiments on TiOCl we applied quasi-hydrostatic pressures up to 16 GPa,
and indeed found large pressure-induced changes in the optical response:
a strong suppression of the transmittance in the infrared and visible range
and a change of the sample color from red to black. In addition, the
near-infrared reflectance abruptly increases at around 12 \nolinebreak GPa.
These findings suggest the occurrence of a pressure-induced
insulator-to-metal transition at 12 GPa.


The crystal structure of TiOCl consists of strongly distorted [TiO$_4$Cl$_2$] 
octahedra.\cite{Schaefer58} A different view of the structure is that of
buckled Ti-O bilayers parallel to the $ab$-plane and separated by Cl ions.
Single crystals of TiOCl were synthesized by chemical vapor
transport from TiCl$_3$ and TiO$_2$.\cite{Schaefer58} The crystal
quality was validated by x-ray diffraction, specific heat, and
magnetic susceptibility measurements.

A diamond anvil cell (DAC) was used for the generation of pressures up to 16 GPa.
Finely ground CsI powder was chosen as quasi-hydrostatic pressure transmitting medium.
For each transmittance and reflectance measurement a small piece (about 80 $\mu$$m$ $\times$
80 $\mu$$m$) was cut from single crystals with a thickness of $\leq$5 $\mu$$m$ and
placed in the hole of a steel gasket. The pressure was determined
by the ruby luminescence method.\cite{Mao86}

\begin{figure}[t]
\includegraphics[width=0.95\columnwidth]{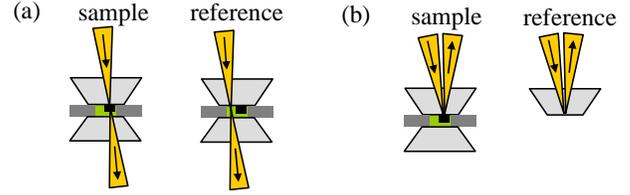}\\
\caption{(Color online) Geometry for high-pressure (a)
transmittance and (b) reflectance measurements of the sample in
the diamond anvil cell.} \label{geometry}
\end{figure}

Pressure-dependent transmittance and reflectance experi\-ments were
conducted at room temperature using a Bruker IFS 66v/S FT-IR
spectrometer with an infrared microscope (Bruker IRscope II). The
reprodu\-cibility of the results was ensured by several
experimental runs on different pieces of eight crystals. For all
measurements the sample was in direct contact with the diamond
anvil on the upper side of the DAC, facing the incoming beam (see
measurement geometry illustrated in Fig.\ \ref{geometry}). In all
other directions the sample was surrounded by the pressure
transmitting medium.

The pressure-dependent transmittance was studied in a wide frequency range
(2000 - 22000 \nolinebreak cm$^{-1}$) for the polarization directions {\bf E}$||$$a,b$.
We measured the intensity I$_{\rm s-CsI}$($\omega$) of the radiation
transmitting the sample [see Fig.\ \ref{geometry}(a)];
as reference, for each pressure we focused the incident radiation spot on the empty
space in the gasket hole next to the sample and obtained the transmitted intensity
I$_{\rm CsI}$($\omega$). The ratio $T$($\omega$)=I$_{\rm s-CsI}$($\omega$)/I$_{\rm CsI}$($\omega$)
is a measure of the transmittance of
the sample, and the corresponding absorbance is calculated according to $A$=log$_{10}$(1/$T$);
$A$ is a measure of the optical conductivity.

\begin{figure}[t]
\includegraphics[width=0.95\columnwidth]{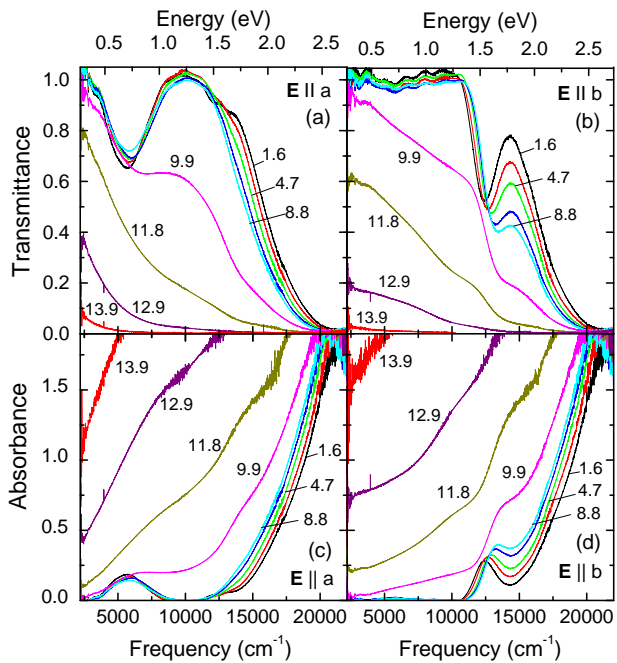}\\
\caption{(Color online) Room-temperature transmittance $T$($\omega$)=I$_{s-CsI}$($\omega$)/I$_{CsI}$($\omega$)
(see text for definitions) of TiOCl as a function of pressure for the polarization (a)
{\bf E}$||$$a$ and (b) {\bf E}$||$$b$. The lower graphs show the corresponding absorbance
$A$=log$_{10}$(1/$T$) as a function of pressure for (c) {\bf E}$||$$a$ and (d) {\bf E}$||$$b$.
The numbers indicate the applied pressures in GPa.
} \label{NIR-VIS}
\end{figure}

Pressure-dependent reflectance measurements were carried out in the frequency range
2000 - 11000 \nolinebreak cm$^{-1}$ for {\bf E}$||$$a,b$.
Reflectance spectra, $R_{\rm s-d}$, of the sample with respect to diamond were
obtained by measuring the intensity $I_{\rm s-dia}$($\omega$) reflected at the
interface between the sample and the diamond anvil [see Fig.\ \ref{geometry}(b)].
As reference, the intensity $I_{\rm dia}$($\omega$) reflected from the inner diamond-air interface
of the empty DAC was used. The reflectance spectra were calculated according to
$R_{\rm s-d}(\omega)=R_{\rm dia}\cdot I_{\rm s-dia}(\omega)/I_{\rm dia}(\omega)$, where
$R_{\rm dia}$ was estimated from
the refractive index of diamond $n_{\rm dia}$ to 0.167 and assumed to be
independent of pressure.\cite{Eremets92,Ruoff94}

\begin{figure}[t]
\includegraphics[width=0.7\columnwidth]{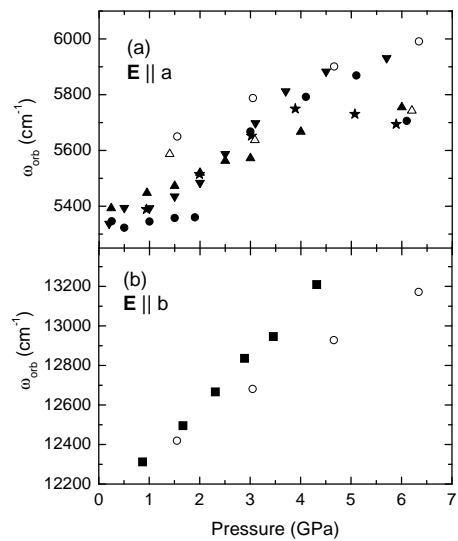}\\
\caption{Pressure-dependent positions of the orbital excitations
of TiOCl, shifting to higher frequencies with increasing pressure,
for (a) {\bf E}$||$$a$ and (b) {\bf E}$||$$b$. Different symbols
correspond to different measurements.} \label{orbital}
\end{figure}

The pressure-dependent transmittance and absorbance spectra of TiOCl for
pressures up to 13.9 \nolinebreak GPa are shown in Fig.\ \ref{NIR-VIS};
above 13.9 GPa the transmitted signal is zero. For the lowest
applied pressure pronounced absorption features are observed at
around 5300 cm$^{-1}$ (0.66 eV) and 12300 \nolinebreak cm$^{-1}$
(1.53 \nolinebreak eV) in the {\bf E}$||$$a$ and {\bf E}$||$$b$
absorbance spectra, respectively [Figs.\ \ref{NIR-VIS}(c) and
\ref{NIR-VIS}(d)]. The two absorption features
can be attributed to excitations between the Ti $3d$ energy levels
whose degeneracy is lifted by the crystal field.\cite{Ruckkamp05a,Ruckkamp05b}
Since our transmittance measurements
were carried out on very thin samples (thickness $\leq$5 $\mu$$m$),
we could determine the precise positions and shapes of the two
orbital excitations. Both absorption peaks are symmetric and
can be described by a Gaussian lineshape.
(For fitting the orbital excitation around 12300 cm$^{-1}$, an
exponential function describing the Urbach tail of the charge
gap\cite{Urbach53} was taken into account.)
The broadening resulting in a Gaussian profile is ascribed to 
vibrational excitations accompanying the electronic 
transitions.\cite{Figgis00}

With increasing pressure the orbital
excitations broaden and continuously shift to higher frequencies
with increasing pressure (see Fig.\ \ref{orbital}). The frequency shifts
indicate an increasing crystal field splitting of the Ti $3d$ levels,
which is most probably due to pressure-induced
changes of the crystal structure, like modifications of the strong
distortions of the [TiO$_4$Cl$_2$] octahedra. However, due to the
lack of crystal structure data under pressure we can only
speculate on this issue.

\begin{figure}[t]
\includegraphics[width=0.97\columnwidth]{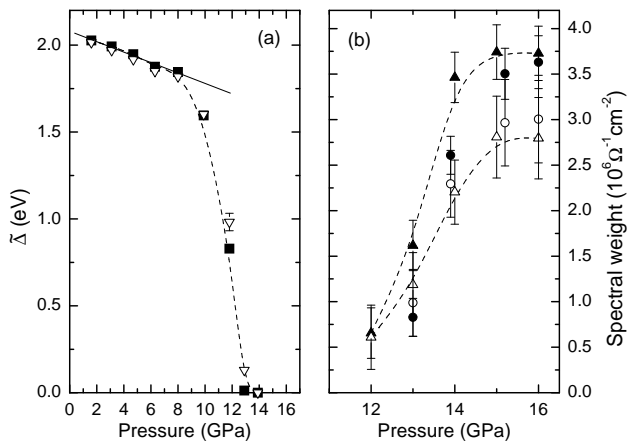}\\
\caption{(a) Charge gap $\tilde{\Delta}$ (see text for definition) as a function of pressure
for {\bf E}$||$$a$ (full symbols) and {\bf E}$||$$b$ (open symbols).
The full line indicates a linear fit to the data; the dashed line
is a guide to the eye.
(b) Pressure dependence of the spectral weight of the optical
conductivity above the insulator-to-metal transition. Full and
open symbols correspond to {\bf E}$||$$a$ and {\bf
E}$||$$b$, respectively.
The same symbols indicate results from the same experimental run.
Dashed lines are guides to the eye.} \label{weight}
\end{figure}

\begin{figure}[b]
\includegraphics[width=0.85\columnwidth]{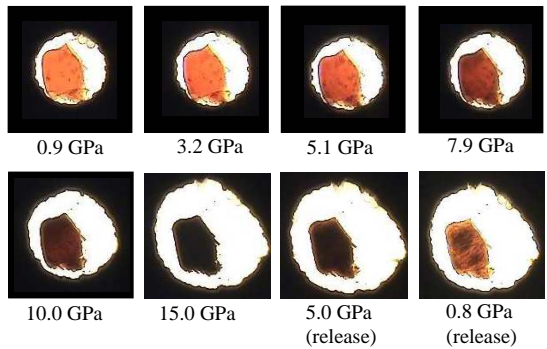}\\
\caption{(Color online) View inside the diamond anvil cell during
the pressure-dependent optical measurements. With increasing
pressure the sample color changes from red to black. Upon pressure
release the sample does not completely recover its original
color.} \label{picture}
\end{figure}

The steep rise of the absorbance above 16000 cm$^{-1}$ ($\approx$2 \nolinebreak eV)
at lowest pressure [see Figs.\ \ref{NIR-VIS} (c) and (d)] can be
attributed to excitations across the charge gap.\cite{Ruckkamp05a,Ruckkamp05b}
We estimated the charge gap, $\tilde{\Delta}$, by a linear extrapolation 
(not shown) of the steep absorption edge .
Starting from the lowest applied pressure, $\tilde{\Delta}$ initially exhibits a
small linear decrease of about 240 \nolinebreak cm$^{-1}$/GPa [see Fig.\ \ref{weight}(a)].
However, above 8~GPa the absorption edge rapidly shifts to lower frequencies
(see also Fig.\ \ref{NIR-VIS}), indicating the abrupt closure of the charge gap.
Above $\approx$12 \nolinebreak GPa the {\bf E}$||$$a,b$ transmittance is
suppressed over the whole studied frequency range, and the overall absorbance 
is strongly enhanced at these high pressures (Fig.\ \ref{NIR-VIS}).

\begin{figure}[t]
\includegraphics[width=0.95\columnwidth]{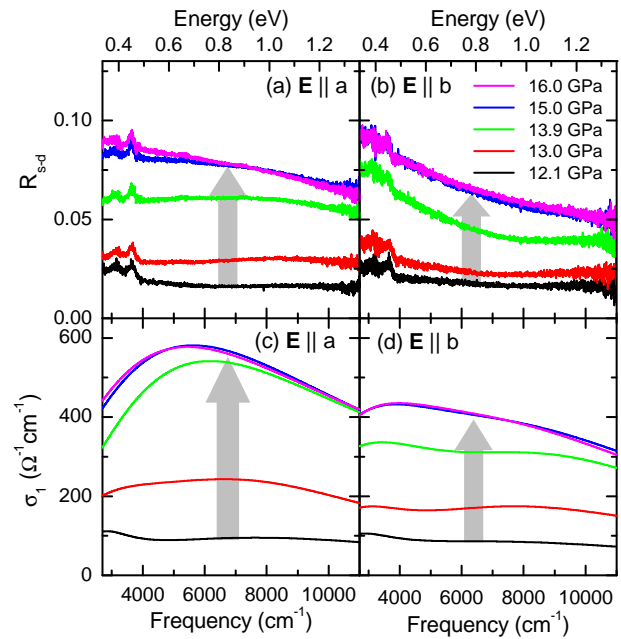}\\
\caption{(Color online) Room-temperature reflectance spectra $R_{\rm s-d}$($\omega$)
as a function of pressure for the polarization (a) {\bf E}$||$$a$ and
(b) {\bf E}$||$$b$.
The lower graphs show the corresponding real part $\sigma_1(\omega)$ of the
optical conductivity obtained by fitting of the $R_{\rm s-d}$($\omega$)
data with the Drude-Lorentz model
for (c) {\bf E}$||$$a$ and (d) {\bf E}$||$$b$. 
Arrows indicate the changes with increasing pressure.}
 \label{reflectance}
\end{figure}

Associated with the rapid reduction of the charge gap
is a change of the sample color. Fig.\ \ref{picture} depicts the
view on the sample inside the DAC at different pressures:
At ambient pressure the sample appears red, since only incident radiation with
frequencies below the charge gap ($\approx$2 \nolinebreak eV) is transmitted. Due to the
strong suppression of the transmittance above $\approx$9 GPa, the sample color changes
from red to black.
Upon pressure release the sample does not completely recover its original color,
as illustrated by some parts of the sample remaining black. It thus seems that the
pressure-induced reduction of the charge gap is not reversible, at least in parts of
the sample.

Since TiOCl becomes opaque
for pressures above $\approx$12~GPa, we completed our spectroscopic study
with pressure-dependent reflectance measurements in the near-infrared range
for {\bf E}$||$$a,b$ [see Figs.\ \ref{reflectance}(a) and (b)].\cite{comment} 
For pressures below 12~GPa the reflectance $R_{\rm s-d}$($\omega$) is very low 
(about 3\%); it cannot be correctly analyzed due to partial transparency of the sample,
and will therefore not be considered.
At around 12 GPa,  $R_{\rm s-d}$($\omega$) abruptly increases
in the whole studied frequency range for both polarization directions.
We ascribe this abrupt increase to additional excitations in the infrared range
induced for pressures above 12 GPa. 
In order to determine their contribution to the optical conductivity of the
sample, we fitted the reflectance spectra with 
the Drude-Lorentz model,\cite{comment1}  as demonstrated in Ref.\ \onlinecite{Kuntscher05}.
The additional excitations were described by a Drude term and two Lorentz oscillators, 
associated with a mid-infrared peak.
The appearance of these additional absorption features is expected
for a filling-controlled or thermally induced Mott transition \cite{Rozenberg96,Imada98}
and was recently also observed in the course of a bandwidth-controlled 
insulator-to-metal transition,\cite{Kezsmarki04} like in the present study.
The real part $\sigma_1(\omega)$ of the optical conductivity obtained from 
the fitting is shown in Figs.~\ref{reflectance}(c) and (d). One clearly 
observes the rapid onset of a broad mid-infrared absorption band above 
12 GPa for both polarizations.

To obtain a measure of the spectral weight of the pressure-induced excitations, 
we integrated the real part $\sigma_1$($\omega$) of the optical conductivity 
between 3000 and 10000 \nolinebreak cm$^{-1}$.
Fig.\ \ref{weight}(b)  shows the so-obtained spectral weight as a function of 
pressure above 12 GPa for both polarizations. It increases approximately linearly
with increasing pressure and appears to saturate above $\approx$15 \nolinebreak GPa.
Together with the concomitant abrupt closure of the charge gap at around 12 GPa 
[see Fig.\ \ref{weight}(a)], this demonstrates the transition-like character 
of the pressure-induced changes in the optical response.
The exact determination of the spectral weight transfer from the charge gap 
excitations to the Drude term and mid-infrared peak requires pressure-dependent 
reflectance data over a broader frequency range; this will be the subject of 
a future study.

In conclusion, we studied the pressure dependence of the
optical response of low-dimensional insulating TiOCl
in the infrared and visible frequency range at room temperature.
The orbital excitations located at ambient pressure at around 0.66
and 1.53 eV for the polarizations {\bf E}$||$$a$ and {\bf
E}$||$$b$, respectively, broaden and shift to higher frequencies
with increasing pressure. The pressure-induced frequency shifts
indicate an increasing crystal field splitting of the Ti $3d$ energy
levels suggestive for crystal structure changes. Both orbital
absorption features are symmetric and have a Gaussian lineshape.
With increasing pressure, a strong suppression of the
transmittance in the infrared and visible energy range and a
change of the sample color are observed. We attribute these
effects to a rapid reduction of the charge gap. 
At $\approx$12 GPa the near-infrared reflectance spectra for both 
studied polarizations abruptly increase. This is attributed to additional
excitations in the infrared frequency range, which can be 
described by a Drude term and a mid-infrared absorption band.
All these findings suggest a pressure-induced insulator-to-metal 
transition in TiOCl at around 12 GPa. \\

We thank M. Dressel for stimulating discussions.
Financial support by the DFG through the Emmy Noether-program and SFB 484 is acknowledged.


\begin{references}
\item[$^*$] Email: kuntscher@pi1.physik.uni- stuttgart.de

\bibitem{Seidel03}
A. Seidel. C. A. Marianetti, F. C. Chou, B. Ceder, and P. A. Lee,
Phys.\ Rev.\ B {\bf 67}, 020405 (2003).

\bibitem{Imai03}
T. Imai and F. C. Chou, cond-mat/0301425 (unpublished).

\bibitem{Kataev03}
V. Kataev, J. Baier, A. M\"oller, L. Jongen, G. Meyer, and A. Freimuth,
Phys.\ Rev.\ B {\bf 68}, 140405 (R) (2003).

\bibitem{Hemberger05}
J. Hemberger, M. Hoinkis, M. Klemm, M. Sing, R. Claessen, S. Horn, and A. Loidl,
Phys.\ Rev.\ B {\bf 72}, 012420 (2005).

\bibitem{Shaz05}
M. Shaz, S. van Smaalen, L. Palatinus, M. Hoinkis, M. Klemm, S. Horn, and R. Claessen,
Phys.\ Rev.\ B {\bf 71}, 100405 (2005).

\bibitem{Ruckkamp05a}
R. R\"uckamp, J. Baier, M. Kriener, M. W. Haverkort, T. Lorenz, G. S. Uhrig, L. Jongen,
A. M\"oller, G. Meyer, and M. Gr\"uninger,
Phys.\ Rev.\ Lett.\ {\bf 95}, 097203 (2005).

\bibitem{Krimmel05}
A. Krimmel et al. (unpublished).

\bibitem{Smaalen05}
S. van Smaalen, L. Palatinus, and A. Sch\"onleber,
Phys.\ Rev.\ B {\bf 72}, 020105(R) (2005).

\bibitem{Saha04}
T. Saha-Dasgupta, R. Valent\'i, H. Rosner, and C. Gros,
Europhys.\ Lett.\ {\bf 67}, 63 (2004).

\bibitem{Lemmens04}
P. Lemmens, K. Y. Choi, G. Caimi, L. Degiorgi, N. N. Kovaleva, A. Seidel, and F. C. Chou,
Phys.\ Rev.\ B {\bf 70}, 134429 (2004).

\bibitem{Ruckkamp05b}
R. R\"uckamp, E. Benckiser, M. W. Haverkort, H. Roth, T. Lorenz, A. Freimuth, L. Jongen, A. M\"oller, G. Meyer,
P. Reutler, B. B\"uchner, A. Revcolevschi, S.-W. Cheong, C. Sekar, G. Krabbes, and M. Gr\"uninger,
New J. Phys.\ {\bf 7}, 144 (2005).

\bibitem{Beynon93}
R. J. Beynon and J. A. Wilson, J. Phys.: Condens.\ Matter {\bf 5}, 1983 (1993).

\bibitem{Craco04}
L. Craco, M. S. Laad, and E. M\"uller-Hartmann,
cond-mat/0410472.

\bibitem{Yamauchi02}
T. Yamauchi, Y. Ueda, and N. M\^ori, Phys.\ Rev.\ Lett.\ {\bf 89}, 057002
(2002).

\bibitem{Imada98}
M. Imada, A. Fujimori, and Y. Tokura,
Rev.\ Mod.\ Phys.\ {\bf 70}, 1039 (1998).

\bibitem{Schaefer58}
H. Sch\"afer, F. Wartenpfuhl, and E. Weise,
Z. Anorg.\ Allg.\ Chem.\ {\bf 295}, 268 (1958).

\bibitem{Mao86}
H. K. Mao, J. Xu, and P. M. Bell, J. Geophys.\ Res.\ [Atmos.] \textbf{91},
4673 (1986).

\bibitem{Eremets92}
M. I. Eremets and Y. A. Timofeev,
Rev.\ Scient.\ Instrum.\ {\bf 63}, 3123 (1992).

\bibitem{Ruoff94}
A. L. Ruoff and K. Ghandehari, in S. C. Schmidt, J. W. Shaner, G. A. Samara, and M. Ross (Eds.),
High Pressure Science and Technology, pp.\ 1523-1525, American Institute of Physics Conference
Proceedings 309, Woodbury, N.Y. (1994).

\bibitem{Urbach53}
F. Urbach, Phys.\ Rev. {\bf 92}, 1324 (1953).

\bibitem{Figgis00}
B. N. Figgis and M. A. Hitchman,
Ligand field theory and its applications,
Wiley-VCH, New York (1996).

\bibitem{comment}
The features around 3200 and 3600 cm$^{-1}$ are artifacts due to multiphonon
absorption by diamond and no intrinsic properties of the studied sample.

\bibitem{comment1}
The background dielectric constant $\epsilon_{\infty}$=3.7 was determined
by a Drude-Lorentz fit of ambient-pressure reflectivity data measured on a 
free-standing thick sample and assumed to be pressure-independent.

\bibitem{Kuntscher05}
C. A. Kuntscher, S. Frank, I. Loa, K. Syassen, T. Yamauchi, and Y. Ueda,
Phys.\ Rev.\ B {\bf 71}, 220502(R) (2005).

\bibitem{Rozenberg96}
M. J. Rozenberg, G. Kotliar, and H. Kajueter,
Phys.\ Rev.\ B {\bf 54}, 8452 (1996).

\bibitem{Kezsmarki04}
I. K\'ezsm\'erki, N. Hanasaki, D. Hashimoto, S. Iguchi, Y. Taguchi, S. Miyasaka, and Y. Tokura,
Phys.\ Rev.\ Lett.\ {\bf 93}, 266401 (2004).


\end{references}
\end{document}